\documentclass[preprint,12pt]{elsarticle}
\usepackage{amsmath}
\usepackage{amssymb}
\usepackage{multirow}
\usepackage{enumerate}
\usepackage{cases}
\usepackage{amsthm}

\usepackage{float}
\usepackage{booktabs}
\usepackage{makecell}
\usepackage{graphicx}
\usepackage{mathrsfs}

\usepackage[misc]{ifsym}
\usepackage{caption}
\usepackage{subfigure}
\usepackage{array}

\makeatletter

\newcommand {\Rmnum} [1] {\expandafter \@slowromancap \romannumeral #1@}
\makeatother

\biboptions{compress}

\date{}

\begin{document}
	
	\begin{frontmatter}
		\title{CNOT-count optimized quantum circuit of the Shor's algorithm}
		
		
		
		\author{Xia Liu}	
		\author{Huan Yang}
		\author{Li Yang\corref{1}}
		\ead{yangli@iie.ac.cn}
		\cortext[1]{Corresponding author.}
		
		\address{State Key Laboratory of Information Security, Institute of Information Engineering, CAS, Beijing, China}
		\address{School of Cyber Security, University of Chinese Academy of Sciences, Beijing, China}
		\address{Institute of Information Engineering, Chinese Academy of Sciences, Beijing, China}

\begin{abstract}

We present improved quantum circuit for modular exponentiation of a constant, which is the most expensive operation in Shor’s algorithm for integer factorization. While previous work mostly focuses on minimizing the number of qubits or the depth of circuit, we try to minimize the number of CNOT gate which primarily determines the running time on a ion trap quantum computer. First, we give the implementation of basic arithmetic with known lowest number of CNOT gate and the construction of improved modular exponentiation of a constant by accumulating intermediate date and windowing technique. Then, we precisely estimate the number of improved quantum circuit to perform Shor’s algorithm for factoring a $n$-bit integer, which is $217\frac{n^3}{\log_2n}+4n^2+n$. According to the number of CNOT gates, we analyze the running time and feasibility of Shor’s algorithm on a ion trap quantum computer. Finally, we discuss the lower bound of CNOT numbers needed to implement Shor's algorithm.

\end{abstract}


\end{frontmatter}


\section{Introduction}
Integer factorization is finding a non-trivial factor of the given compositive number. It is believed to be a hard mathematic problem for which no classical polynomial-time algorithm has yet been discovered. As the most representative and compelling quantum algorithm, Shor’s algorithm[1,2] can factor integers with only polynomial time in theory, which offer an exponential speedup over its classical counterpart(number field sieve[3] up to now). It poses a serious threat to the classical public cryptosystem whose security is based on integer factorization, including RSA[4] which is widely used in key exchange and digital signature.

Quantum computer implements quantum computation which accepts quantum states representing superposition of all different possible inputs and simultaneously evolves them into corresponding outputs by a sequence of quantum gates. Quantum computation can be decribed as a quantum circuit in which the quantum gates represent the unitary transformations. Since the appearance of Shor’s algorithm in 1994, a lot of efforts have been devoted to the design and improvements of its quantum circuit and its improvement in terms of the number of qubits and the depth of circuit. The first work is [5], in which Vedral et al. provided an explicit quantum circuit construction of basic arithmetic operations from addition to modular exponentiation. Beckman et al. [6] estimated the number of qubits and operations required of Shor’s algorithm: a $n$-bit integer can be factored in $O(n^3)$ operations with $5n+1$ qubits. Miquel et al.[7] analyzed the impact of losses and decoherence on the quantum circuit of Shor’s algorithm. In terms of the number of qubits required: Beauregard[10] contructed a quantum circuit of Shor’s algorithm with $2n+3$ qubits with using the QFT-based adder[8] and semiclassical QFT[9]; Takahashi and Kunihiro[11] reduced the number of qubits to $2n+2$ with comparator in modular addtion operation, which is the lowest known number of qubits so far; Häner et al.[12] constructed a quantum circuit of Shor’s algorithm with $2n+2$ qubits as well based on a purely Toffoli-based adder which eliminats most of the cost overheads originating from rotation synthesis and enable testing and debugging. In terms of the depth of circuit: Zalka[13] reducd the depth of circuit of Shor’s algorithm to $2^{17}n^{1.2}$ but required $24n\sim96n$ qubits with FFT-based fast multiplication; Pavlidis and Gizopoulos’s circuit[14] implemented division to compute modular mutiplication, having a depth of $2000n^2$ and requiring $9n+2$ qubits.

In this paper, we consider quantum circuit of Shor’s algorithm with low CNOT-count, that is a completely different perspective from previous work. Clifford+T set is general to approximate an arbitrary unitary transform with arbitrary precision in quantum computation. As the only double-qubit gate in Clifford+T set, the time of the CNOT gate acting on the non-adjacent qubits is much higher than that of other single-qubit gates in the ion trap quantum computer. Furthermore, CNOT gates cannot be parallel in ion trap quantum computer, even if acting on completely unrelated qubits. Therefore, the total CNOT-count in quantum circuit primarily determains the running time of quantum algorithm in the ion trap quantum computer. We improve the quantum circuit of Shor’s algorithm to reduce CNOT-count by applying window technique[15], Montgomery multiplication[16] and pebbling technique[17] to modular exponentiation operation, which is the most computational intensive ingredient of Shor’s algorithm. Besides, we also esimate the time to run Shor’s algorithm once in the ion trap quantum computer and analyze the feasibility of Shor’s algorithm, based on the CNOT-count of our improved circuit and the lower time limit of CNOT gate in the ion trap quantum computer in [18]. 
   
The rest of this paper is organized as follows. Section 2 describes Shor’s algorithm and the basic arithmetic circuits constructed previously. Section 3 is our works on the arithmetic circuits and the circuit of modular exponentiation operation as well as analysis of corresponding CNOT-count. Section 4 is about the lower bound of CNOT gate needed to run Shor's algorithm. The last part is the result about the time estimation and feasibility of Shor’s algorithm. In the figures of this paper, we use the black triangles on the right side of gate symbols to indicate quantum registers which are modified and holding the result of the computation.

\section{Preliminaries}
\subsection{Shor's algorithm}
Given the integer $N$ to be factored, Shor’s algorithm consists of the quantum order-finding and the classical post-processing. Let $a$ be a randomly chosen integer which is less than $N$ and coprime to $N$, the order of $a$ is the least positive integer $r$ such that $a^r\equiv1\mod N$. The quantum order-finding shown in Figure 1 requires two work quantum registers, The first quantum register consists of $2n$ qubits which is set to $|0\rangle$ initially and the second quantum register consists of $n$ qubits which is set to $|1\rangle$ initially, where $n=\lceil\log N\rceil$ is the number of bits to represent $N$. As shown in figure 1, there are four steps in the quantum order-finding:
\begin{enumerate}[i]
	\item Apply Hadamard transform $H^{\otimes2n}$ to the first quantum register, create a superposition state in which the elements correspond to the exponents in step ii:
	\[\frac{1}{2^n}\sum_{x=0}^{2^{2n}-1}|x\rangle|1\rangle;\] 	
	\item Compute the modular exponentiation by the constant $a$:
    \[\frac{1}{2^n}\sum_{x=0}^{2^{2n}-1}|x\rangle|a^x\mod N\rangle;\]	
	\item 	Apply the quantum Fourier transform $QFT_{2^{2n}}$ to the first quantum register:
	\[\frac{1}{2^{2n}}\sum_{x=0}^{2^{2n}-1}\sum_{y=0}^{2^{2n}-1}\exp{\frac{2\pi ixy}{2^{2n}}}|y\rangle|a^x\mod N\rangle;\]	
	\item Measure the first quantum register and find the order of $a$ with high probability by classical post-processing to the measured data.
\end{enumerate}
\begin{figure}[H]
	\centering
	\includegraphics[scale=0.8]{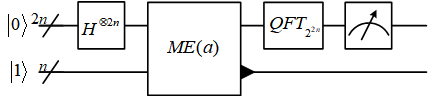}
	\caption{Overall Shor's algorithm for interger factorization.}
	\label{fig:1}
\end{figure}

The most expensive operation in the quantum order-finding is the modular exponentiation by the classical known constant $a$ in step ii, which is denoted as $ME(a)$ in figure 1. $ME(a)$ applies the following transform to its input quantum states:
\[|x\rangle|1\rangle\rightarrow|x\rangle|a^x\mod N\rangle\]

According to the binary expansion of $x$:
\[x=2^0x_0+2^1x_1+\cdots+2^{2n-1}x_{2n-1}\]

Notice that $a^x$ can be written as :
\[a^x=a^{2^0x_0}\cdot a^{2^1x_1}\cdots a^{2^{2n-1}x_{2n-1}}\]

Therefore, starting from $|1\rangle$, the computation of $a^x\mod N$ can be decomposed into $2n$ modular multiplications by the classical known constant $a^{2^i}\mod N$ controlled by the corresponding qubit $|x_i \rangle$ where $i$ takes value between $0$ and $2n-1$. 
\begin{figure}[H]
	\centering
	\includegraphics[scale=0.6]{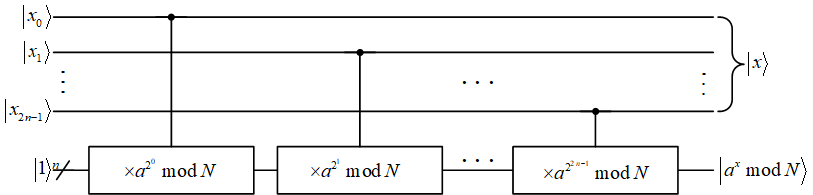}
	\caption{Schematic diagram of the compution of $a^x\mod N$}
	\label{fig:2}
\end{figure}
Similarly, the product $ax$ which multiplies the input quantum state $|x\rangle$ by the classical known constant $a$ can be written as:
\[ax=2^0ax_0+2^1 ax_1+\cdots+2^{n-1} ax_{n-1}\]

Therefore, starting from $|0\rangle$, the computation of $ax\mod N$ can be decomposed into $n$ modular additions by the classical known constant $a{2^i}\mod N$ controlled by the corresponding qubit $|x_i\rangle$ where $i$ takes value between $0$ and $2n-1$.
\begin{figure}[H]
	\centering
	\includegraphics[scale=0.7]{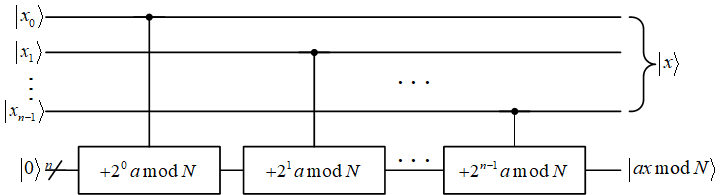}
	\caption{Schematic diagram of the computation of $ax\mod N$}
	\label{fig:3}
\end{figure}
We denoted the modular multiplication operation by the classical known constant $a$ as $MM(a)$, which performs the following transform to its input quantum states:
\[|x\rangle |0\rangle\rightarrow|x\rangle|ax\mod N\rangle\]

Since $MM(\cdot)$ operation multiplies the input quantum state $|x\rangle$ by the classical known constant to a different quantum register which is initially set to $|0\rangle$, the direct approach to compute $a^x\mod N$ with $2n$ $MM(\cdot)$ operations will accumulate the intermediate data of each $MM(a^{2^i })$ operation. The following method proposed by Vedral et al.[5] is widely adopted to implement the in-place modular multiplication to the input quantum state $|x\rangle$ by the classical known constant $a$:
\begin{enumerate}[i]
	\item Apply $MM(a)$ to the input quantum register and the ancillary quantum register initially set to $|0\rangle$:
	\[|x\rangle|0\rangle\rightarrow|x\rangle|ax\mod N\rangle;\] 	
	\item Swap the quantum states of the the input quantum register and the ancillary quantum register:
	\[|ax\mod N\rangle|x\rangle;\]	
	\item Apply $MM^{-1} (a^{-1})$ to the input quantum register and the ancillary quantum register:
	\[|ax\mod N\rangle|0\rangle.\]
\end{enumerate}

Therefore, as shown in figure 4, $ME(a)$ operation can be constructed by $MM(\cdot)$ operations. 
\begin{figure}[H]
	\centering
	\includegraphics[scale=0.7]{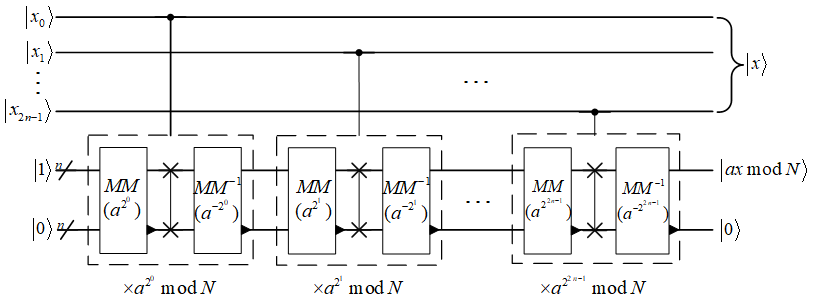}
	\caption{The construction of $ME(a)$ operation by $MM(\cdot)$ operations}
	\label{fig:4}
\end{figure}

\subsection{Previous works on basic arithmetic}
We compared the previous different types of basic arithmetic circuits and selected the circuits used to construct the Shor's algorithm with the fewest CNOT gates required.

\textbf{Addition.} We compared different types of addition circuits and found that Cuccaro et al's addition[8](hereinafter the CDK adder) is the lowest known in the CNOT-count. Based on the standard decomposition of Toffoli gate into Clifford+T set which contains $6$ CNOT gates[19], the CNOT-count of $CDK$ adder for $n$-bit binary integers is $16n+1$.

\textbf{Addition by a constant.} Addition by a constant $a$ can be constructed from $CDK$ adder, as shown in figure 5. First bind the constant $a$ to an input quantum register of $CDK$ adder initially in $|0\rangle$, then apply $CDK$ adder to compute the sum, finally recover the input quantum register to $|0\rangle$ by the same way as first step. The binding operation of a constant $a$ is to apply $X$ gates to the appropriate qubits which are corresponding to 1 in the binary representation of $a$. Different form addition with $2$ unknown addicands both in the form of quantum state, here the adder can be simplified by the known constant. Therefore, the CNOT-count of addition by a constant for $n$-bit binary integers is $13n+1$. 
\begin{figure}[H]
	\centering
	\includegraphics[scale=0.6]{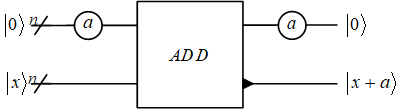}
	\caption{Addition by a constant $a$ constructed from $CDK$ adder}
	\label{fig:7}
\end{figure}
\textbf{Controlled addition.} We use controlled $CDK$ adder given in [21] by controlling all the $MAJ$ and $UMA$ blocks of CDK adder. The CNOT-counts of the controlled $MAJ$ block and controlled $UMA$ block are both $13$, so that the CNOT-count of controlled $CDK$ adder for $n$-bit integers is $26n+6$.

\textbf{Controlled addition by a constant.} Based on the addition by a constant, we control the two binding operations of the constant with the intermediate uncontrolled $CDK$ adder as shown in the left of figure 6 instead of controlled $CDK$ adder only as shown in the right of figure 6. Considering that CNOT-count of binding operation of a constant is $\frac{1}{2}$ in average, the CNOT-count of the controlled addition for $n$-bit integers by a constant is $17n+1$.
\begin{figure}[H]
	\centering
	\includegraphics[scale=0.6]{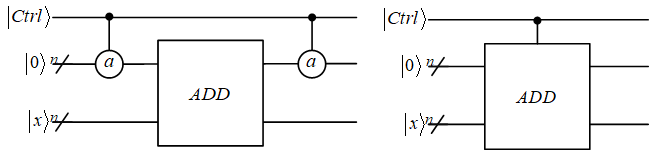}
	\caption{Controlled addition by a constant based on the addition by a constant.}
	\label{fig:8}
\end{figure}
\textbf{Comparison.} We use the comparison based on the $MAJ$ blocks given in [21], of which the CNOT-count is $16n+1$ for $n$-bit integers. 
\textbf{Comparison by a constant.} We use the comparison by a constant given in [20] as well, which is similar to the construction of addtion by a constant. And the CNOT-count of comparison by a constant for $n$-bit binary integers is $12n+1$.

\section{Methodology}
\subsection{Improvement of basic arithmetic}
Based on the basic arithmetic circuits in the previous section, we improve the modular addition, shift and modular doubling circuits. At the same time, we construct the controlled comparison and controlled modular addition circuits according to the previous comparison and modular addition circuits.

\textbf{Controlled comparison.} As shown in figure 7, we construct the controlled comparison by controling the CNOT gate and $X$ gate on the qubit holding the result of comparison. The CNOT-count of controlled comparison for $n$-bit integers is $16n+7$.
\begin{figure}[H]
	\centering
	\includegraphics[scale=0.6]{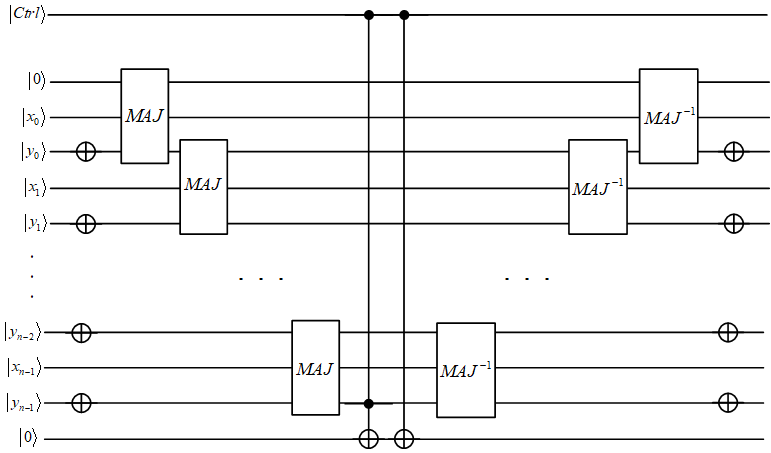}
	\caption{Controlled comparison}
	\label{fig:10}
\end{figure}
\textbf{Modular addition.} We improve the modular addition as show in figure 8, in which the substraction is the inverse of addition. The CNOT-count of our modular addition is $61n+16$.
\begin{figure}[H]
	\centering
	\includegraphics[scale=0.6]{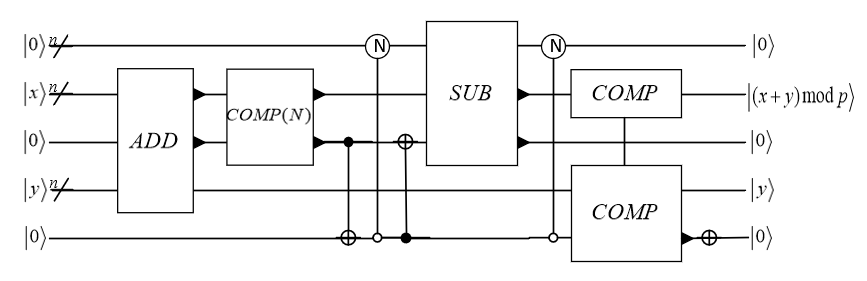}
	\caption{Modular addition}
	\label{fig:11}
\end{figure}
\textbf{Controlled modular addition.} We construct the controlled modular addition by controlled the first addition and the last comparison of modular addition as show in figure 9. The CNOT-count of our controlled modular addition is $71n+27$.
\begin{figure}[H]
	\centering
	\includegraphics[scale=0.6]{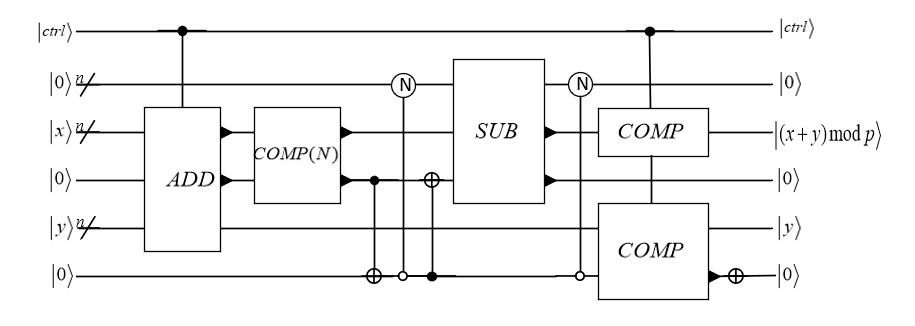}
	\caption{Controlled modular addition}
	\label{fig:12}
\end{figure}
\textbf{Shift.} Note that there is no need for swaping two qubit with SWAP operation if a qubit is known in the state of $|0\rangle$, so that we contruct the left shift and right shift for a $n$-qubit quantum register as shown in figure 10, of which the CNOT-count are both $2n$.
\begin{figure}[H]
	\centering
	\includegraphics[scale=0.6]{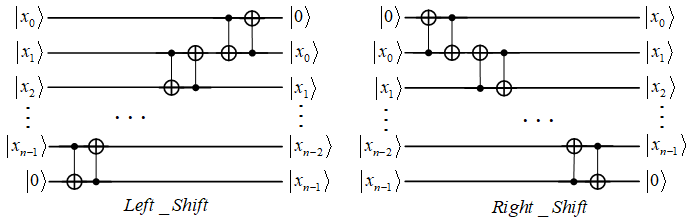}
	\caption{Left shift and right shift}
	\label{fig:13}
\end{figure}
\textbf{Modular Doubling.} As shown in figure 11, We improve the modular doubling which replace the substruction of the constant $N$ in [22] with the comparison of the constant $N$. The CNOT-count of our modular doubling is $31n+15$.
\begin{figure}[H]
	\centering
	\includegraphics[scale=0.7]{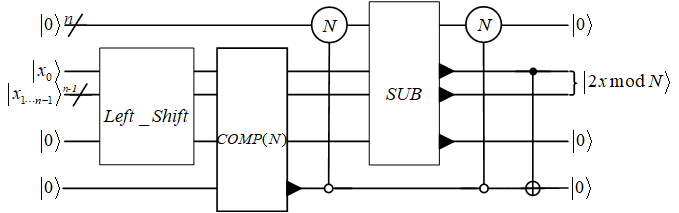}
	\caption{Modular doubling}
	\label{fig:14}
\end{figure}

\subsection{A new circuit running shor's algorithm}
\textbf{Accumulate the intermediate data.} As shown in figure 12, instead of erasing the intermediate data in each controlled $MM(\cdot)$ operation described as Vedral et al.’s method, we accumulate the intermediate data until the modular exponentiation $a^x$ is computed by the last controlled $MM(a^{2^{2n-1}}\mod N)$ operation and erase the intermediate data is by the controlled inverse $MM(\cdot)$ operaions in reverse order except for the last controlled $MM(a^{2^{2n-1}}\mod N)$ operation.
\begin{figure}[H]
	\centering
	\includegraphics[scale=0.4]{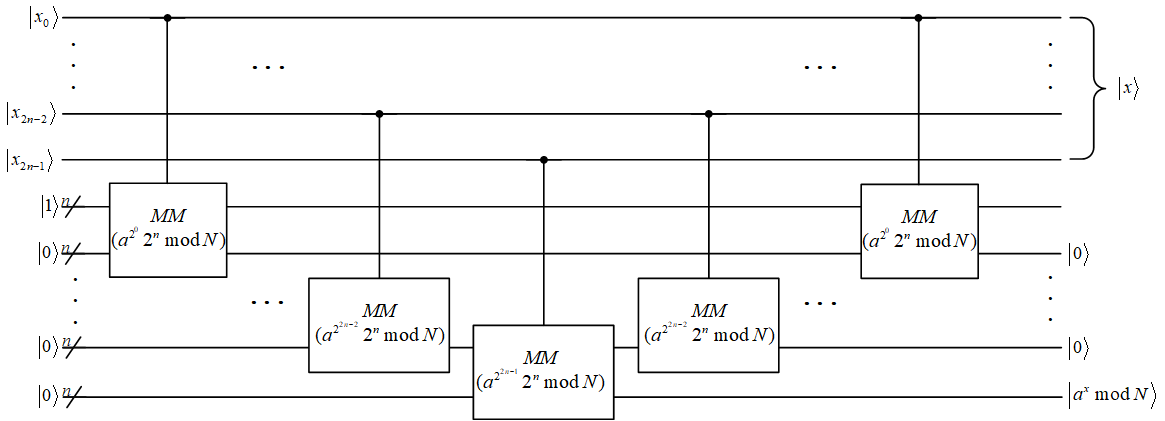}
	\caption{The construction of $ME(a)$ opertion by $MM(\cdot)$ operations with the intermediate date accumulated. }
	\label{fig:15}
\end{figure}
\textbf{Windowing technique.} Windowing technique is widely used to reduce the number of operation in classical computation, such as the fast implementation of CRC parity check [23]. Gidney[15] showed that it is also useful to optimize quantum circuits in quantum computation and presented various windowed quantum arithmetic circuits, including a windowed modular exponentiation with nested windowed modular multiplications. The key of windowing technique in quantum computation is to merge several controlled operation acting on the target quantum state into a single operation acting on the target quantum state and a corresponding special quantum state which is created and recovered by table lookup operation. 

Since $ME(a)$ operation can be decomposed into a series of controlled $MM(\cdot)$ operations, it is suitable to apply windowing technique to $ME(a)$ operation. We iterate all the control qubits in groups with the window size $m$ instead of individually. For the $m$ controlled $MM(\cdot)$ operations in each group, we merge them as the following steps: 
\begin{enumerate}[i]
	\item Retrieve the value which the result is actually multiplicated by from the precomputed table by the $m$ control qubits. Create the special quantum state corresponding to the value found in the ancillary quantum register.
	\item Modular multiply the value of the target quantum state by the value of the special quantum state.	
	\item Retrieve the value which the result is actually multiplicated by from the precomputed table by the $m$ control qubits. Recover the special quantum state corresponding to the value found in the ancillary quantum register.
\end{enumerate}

The table lookup operation in step i and iii performs $|x\rangle|0\rangle\rightarrow|x\rangle|T_x\rangle$, where $T_x$ is the value found from the classical precomputed table addressed by $x$. We give the quantum circuit of table lookup operation without control qubit based on [15] showed in Figure 13. 
\begin{figure}[H]
	\centering
	\includegraphics[scale=0.6]{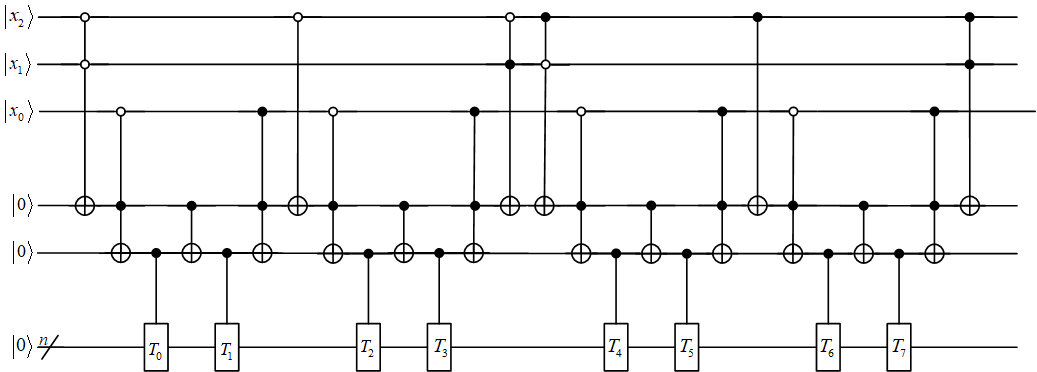}
	\caption{Quantum circuit of table lookup operation in the case where the window size $m=3$. If the control qubits contains the value $i$, then bind the value found $T_i$ from the classical precomputed table addressed by $i$ to the lowest register by applying the $X$ gate to the appropriate qubits of the lowest register. The binding operation is denoted as a circle with the value to bind in the figure. }
	\label{fig:16}
\end{figure}
The modular multiplication with the two factors both in the form of quantum state can be computed by modular multiplication operation denoted as $MM$, which performs the following transform to its input quantum states:
\[|x\rangle|y\rangle|0\rangle\rightarrow|x\rangle|y\rangle|x\cdot y\mod N\rangle.\]

The quantum circuit of windowed $ME(a)$ operation based on the construction with the intermediate date accumulated is shown in figure 14.

\textbf{Modular multiplication.} Roetteler et al.[22] showed the quantum circuits of two approaches to compute modular multiplication of two factors both in the form of quantum state: Fast modular multiplication[24] and Montgomery modular multiplication[16].

By the binary expansion of the first factor $x$, the product $x\cdot y$ can be written as:
\[x\cdot y=\sum_{i=0}^{n-1}2^ix_iy=x_0y+2(x_1y+2(x_2y+\cdots+2x_{n-1}y\cdots))\]

So fast modular multiplication decompose $x\cdot y\mod N$ into a sequence of conditional modular additions and modular doublings. Based our constructions of basic arithmetic, we improve the circuit of fast modular multiplication which is shown in figure 15 and the CNOT-count is $102n^2-54n-42$.
\begin{figure}[H]
	\centering
	\includegraphics[scale=0.4]{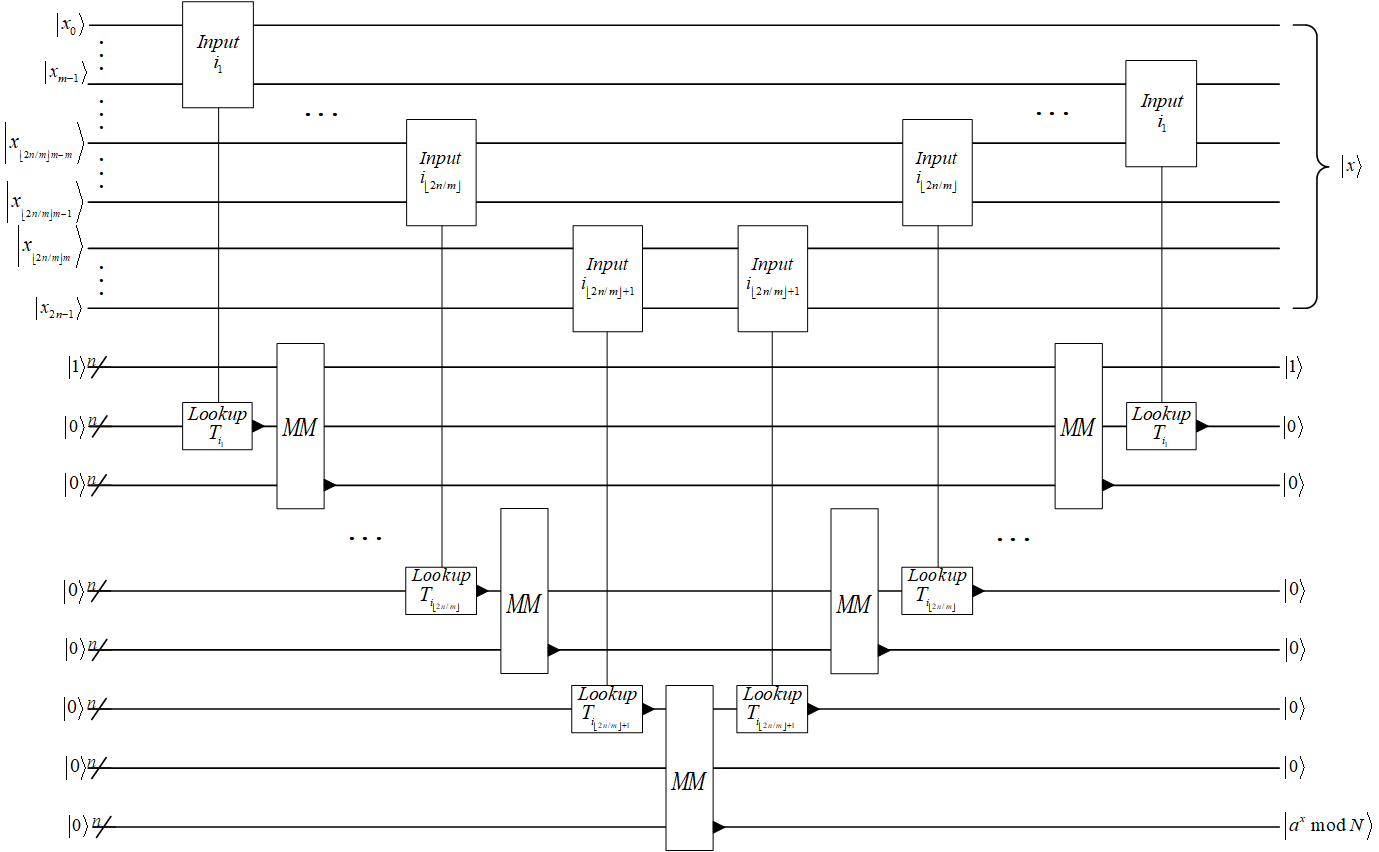}
	\caption{Windowed $ME(a)$ operation based on the construction with the intermediate date accumulated.}
	\label{fig:17}
\end{figure}
\begin{figure}[H]
	\centering
	\includegraphics[scale=0.6]{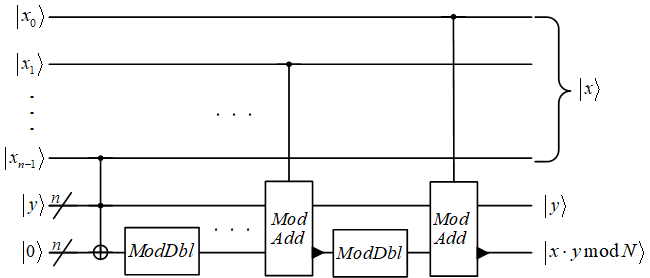}
	\caption{Fast modular multiplication. We replace the first controlled modular addition with a controlled copying operation since the modular sum equals to another addicand if an addicand is $0$.}
	\label{fig:18}
	\end{figure}
Montgomery modular multiplication computes $\frac{x\cdot y}{2^n}\mod N$ instead of $x\cdot y\mod N$ for the input factors $x$ and $y$ in the form of quantum state. Figure 16 gives the whole quantum circuit of Montgomery modular multiplication, which consists of the forward Montgomery modular multiplication and the backward Montgomery modular multiplication. If an input factor $x$ is in the form of Montgomery representation $(x\cdot2^n\mod N)$, then the result of the Montgomery modular multiplication will be $x\cdot y\mod N$, which is what we require actually. 
\begin{figure}[H]
	\centering
	\includegraphics[scale=0.4]{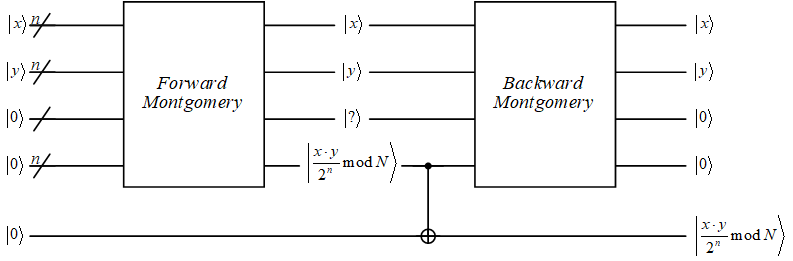}
	\caption{Whole quantum circuit of Montgomery modular multiplication which performs $|x\rangle|y\rangle|0\rangle\rightarrow|x\rangle|y\rangle|\frac{x\cdot y}{2^n}\mod N\rangle$. The forward Montgomery modular multiplication computes $\frac{x\cdot y}{2^n}\mod N$ with the information of intermediate result in each round held in ancillary qubits. To recover the ancillary qubits of the forward Montgomery modular multiplication, one can copy the result $\frac{x\cdot y}{2^n}\mod N$ to another new quantum register and apply the backward Montgomery modular multiplication, that is to run the circuit of forward Montgomery modular multiplication backwards. }
	\label{fig:19}
\end{figure}
Based our constructions of basic arithmetic, the CNOT-count of the forward Montgomery modular multiplication which is shown in figure 17 is $45n^2+17n+8$, while the CNOT-count of the whole Montgomery modular multiplication is $90n^2+35n+16$.
\begin{figure}[H]
	\centering
	\includegraphics[scale=0.4]{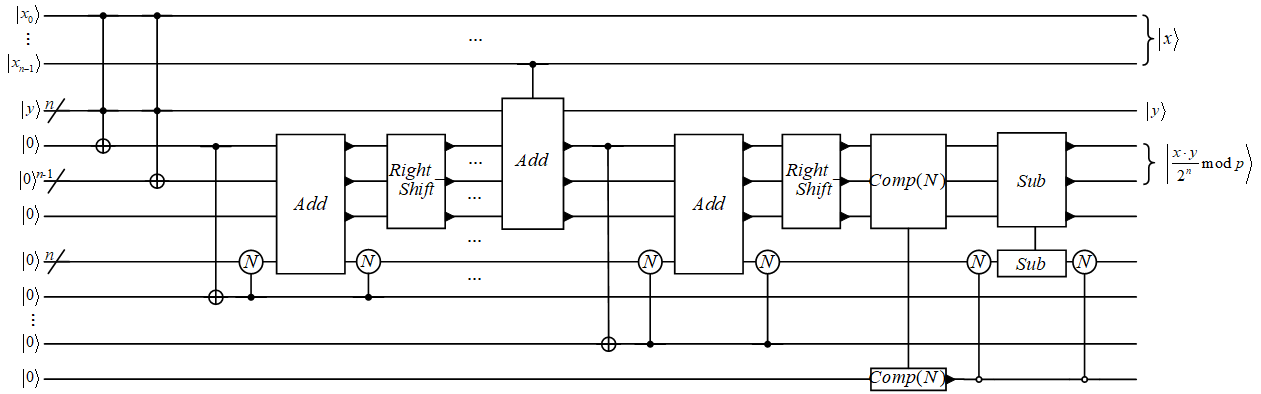}
	\caption{Forward Montgomery modular multiplication. We replace the first controlled modular addition with a controlled copying operation since the sum equals to another addicand if an addicand is $0$.}
	\label{fig:20}
\end{figure}

\section{Discussion and conclusion}

Given the $n$-bit integer to be factored and the window size $m$, the precise number of CNOT gate in our quantum circuit for modular exponentiation is $(\lceil\frac{2n}{m}\rceil-1)[(n+13) 2^m+90n^2+34n-10]+(n+13)2^{2n-m(\lceil\frac{2n}{m}\rceil-1)}+102n^2-54n-42$, which achieves the minimum related to the window size $m$ for the input bit size $n$. We fit the data with a range of input bit size and get the result of $217\frac{n^3}{\log_2n}$. Combine the $4n^2+n$ CNOT gates used for QFT on $2n$ qubits, We conclue that the total number of CNOT gate ro run Shor’s algorithm once is $217\frac{n^3}{\log_2n}+4n^2+n$. [18] gives the lower limit of time for executing a CNOT gate in an ion trap quantum computer, which is about $2.85\times10^{-4}s$. Combined with the number of CNOT to run Shor's algorithm, the time  to break $1024$-$bit$ RSA is at least 72 years after three levels of coding.

If we assume that the time required to run the Shor's algorithm is $T$, the time required to execute a CNOT is $t$ and the lower bound of the CNOT gate is $N$, which is a function of the number of qubits $n$. So the lower bound of running Shor's algorithm can be expressed as $T=N(n)t$. Modular exponentiation can be constructed by modular multiplication and modular multiplication can be constructed by modular addition. So the number of CNOT gates required for modular exponentiation must be greater than modular addition. Since the quantum circuit of modular addition adds the modular operation, the CNOT number required by modular addition must be larger than that of addition circuit. For two $n$ qubits $x, y$, there has 
\begin{align*}
c_{i+1}&=x_ic_i+y_ic_i+x_iy_i\\
&=x_i+(x_i+y_i)(x_i+c_i),
\end{align*}
\[
	s_i=x_i+y_i+c_i,
\]
where $x_i$ and $y_i$ are the $i$-$th$ bit of the binary representation of $x,y$, $c_{i+1}$ is the $i$-$th$ carry, and $s_i$ is the sum of the $i$-$th$ bit. Therefore, each qubit addition requires at least 1 Toffoli and 3 CNOTs. So the addition of $n$ qubits requires at least $9n$ CNOTs. Thus, we can get that the lower bound of the number of CNOT gates required to run Shor's algorithm is $9\frac{n^3}{\log_2n}$. The range of this results is a bit large, and we shoule calculate the lower bound more accurately in our next work.

In this paper, we improve the quantum circuit of basic arithmetic including addition, controlled addition and comparison. We construct the circuit of controlled comparison, controlled modular addition and modular multiplication. Based on these work, the quantum circuit running Shor's algorithm is improved, and we calculate the number of CNOT gates required by the algorithm. The time required by Shor's algorithm to attack $1024$-bits RSA is estimated. Finally, the lower bound of the required CNOT gate independent of algorithm improvement is discussed.


\end{document}